\documentclass[11pt]{article}
\usepackage{natbib}
\usepackage{latexsym}
\usepackage{amsfonts}
\usepackage{amsmath}
\usepackage{amssymb}

\makeatletter%
\def\nottoobig#1{{\hbox{$\left#1\vcenter to1.111\ht\strutbox{}\right.\n@space$}}}
\makeatother%

\makeatletter %

\newlength{\filength}
\newsavebox{\gcbox}
\sbox{\gcbox}{\framebox[\filength]{\rule{0ex}{2ex}}}

\newcommand{\qedblob}{\mbox{\rule[-1.5pt]{5pt}{10.5pt}}}
\def\literalqed{{\ \nolinebreak\hfill\mbox{\qedblob\quad}}}

\def\qed{\literalqed}

\makeatother%

\newcommand{\naturalnumber}{\ensuremath{{  \mathbb{N} }}}

\newcommand{\sharpp}{{\rm \#P}}

\newcommand{\sat}{{\rm SAT}}

\newcommand{\parityp}{{\rm \oplus P}}
\newcommand{\up}{{\rm UP}}

\newcommand{\fewp}{{\rm FewP}}
\newcommand{\coup}{{\rm coUP}}

\newcommand{\p}{{\rm P}}
\newcommand{\littlep}{{\rm p}}

\newcommand{\np}{{\rm NP}}

\newcommand{\conp}{{\rm coNP}}

\newcommand{\ph}{\ensuremath{{\rm PH}}}
\newcommand{\few}{{\ensuremath{\rm Few}}}
\newcommand{\const}{{\ensuremath{\rm Const}}}
\newcommand{\littleconst}{{\ensuremath{\rm const}}}

\newtheorem{thm}{Theorem}[section]
\newtheorem{cor}[thm]{Corollary}

\newtheorem{prop}[thm]{Proposition}
\newtheorem{defn}[thm]{Definition}

\newcommand{\pairs}[1]{\mathopen\langle{#1}\mathclose\rangle}

\newcommand{\manyone}{\ensuremath{\,\leq_{\rm m}^{{\littlep}}\,}}

\newcommand{\sigmastar}{\ensuremath{\Sigma^\ast}}

\newcommand{\calc}{\ensuremath{{\mathbb{C}}}}
\newcommand{\cala}{\ensuremath{{\mathbb{A}}}}

\newcommand{\bigo}{{O}}

\newcommand{\condition}{\,\nottoobig{|}\:}
\newcommand{\conditionX}{\,\nottoobig{|}}
 
\def\land{{\; \wedge \; \allowbreak}}

\newcommand\integers{{\mathbb{Z}}}
\def\Z{\integers}
\newcommand{\counting}{\mbox{\tt Counting}}
\newcommand{\seq}{\subseteq}
\newcommand{\spp}{\mbox{\rm SPP}}
\newcommand{\fp}{{\rm FP}}
\def\nats{\naturalnumber}
\newcommand{\bh}{{\rm BH}}

\newenvironment{construction}{\bigbreak\begin{block}}{\end{block}
    \bigbreak}

\newenvironment{block}{\begin{list}{\hbox{}}{\leftmargin 1em
    \itemindent -1em \topsep 0pt \itemsep 0pt \partopsep 0pt}}{\end{list}}

\begin{document}
\bibliographystyle{plain}

\title{A Second Step Towards Complexity-Theoretic Analogs of Rice's Theorem}

\author{
Lane A. Hemaspaandra\thanks{%
Email: {\tt lane@cs.rochester.edu}.
Supported in part 
by grants NSF-CCR-9322513 and 
NSF-INT-9513368/DAAD-315-PRO-fo-ab.  Work done in part while 
visiting 
Friedrich-Schiller-Universit\"at Jena.}
\\
Department of Computer Science\\
University of Rochester\\
Rochester, NY 14627, USA\\
\and 
J\"{o}rg Rothe\thanks{%
Email: {\tt rothe@informatik.uni-jena.de}.
Supported in part 
by grant
NSF-INT-9513368/DAAD-315-PRO-fo-ab.
Work done in part during a postdoctoral year at 
the University of Rochester, supported by a 
NATO Postdoctoral Science Fellowship
from the Deut\-scher Aka\-de\-mi\-scher Aus\-tausch\-dienst
(``Ge\-mein\-sames Hoch\-schul\-sonder\-pro\-gramm~III 
von Bund und L\"andern'').}
\\
Institut f\"ur Informatik\\
Friedrich-Schiller-Universit\"at Jena\\
07740 Jena, Germany\\
}

\date{}

\maketitle 

\begin{abstract}
    Rice's Theorem states that every nontrivial language property
    of the recursively enumerable sets 
    is
    undecidable. Borchert and Stephan~\cite{bor-ste:c:rice} 
    initiated the search for 
    complexity-theoretic analogs of Rice's Theorem. In particular, they
    proved that every nontrivial counting property of
    circuits is UP-hard, and that a number of closely related 
    problems are SPP-hard.  

    The present paper studies whether their UP-hardness
    result itself can be improved to SPP-hardness.
    We show that their UP-hardness result cannot be
    strengthened to SPP-hardness unless unlikely complexity class
    containments hold. Nonetheless, we prove that every P-constructibly
    bi-infinite counting property of circuits is
    SPP-hard.  
    We also raise their general lower bound
    from unambiguous nondeterminism to constant-ambiguity
    nondeterminism.
\end{abstract}

\section{Introduction}
Rice's 
Theorem~(\cite{ric:j:nontrivial-properties-of-re-sets,ric:j:re-sets-and-their-key-arrays},
see~\cite{bor-ste:c:rice})
states that every nontrivial language 
property of the recursively enumerable sets
is either RE-hard or coRE-hard---and thus is certainly 
undecidable (a corollary that itself is often
referred to as Rice's Theorem).

\begin{thm}[Rice's Theorem, Version I] \quad
  \label{t:rversiontwo-nowone}
  Let ${\cala}$ be a nonempty proper subset of the class of
  recursively enumerable sets. Then either the halting problem or its
  complement many-one reduces to the problem: Given a Turing
  machine~$M$, is $L(M) \in {\cala}$?
\end{thm} 

\begin{cor}[Rice's Theorem, Version II]
  \label{cwast:rversionone-nowtwo} \quad
  Let ${\cala}$ be a nonempty proper subset of the class of
  recursively enumerable sets. Then the following problem is
  undecidable: Given a Turing machine~$M$,
  is $L(M) \in {\cala}$?
\end{cor}

Rice's Theorem conveys quite a bit of information about the nature of
programs and their semantics. Programs are completely nontransparent.
One can (in general) decide {\em nothing\/}---emptiness, nonemptiness,
infiniteness, etc.---about the languages of given programs
other than the trivial fact that each accepts some language
and that language is a 
recursively enumerable language.\footnote{One must stress 
that Rice's Theorem refers to the languages 
accepted by the
programs (Turing machines)
rather than to machine-based actions of the programs (Turing
machines)---such
as whether they run for at least seven steps on input 1776 (which is
decidable) or whether for some input they do not halt
(which is not decidable, but
Rice's Theorem does not speak directly to this issue, that is,
Rice's Theorem does not address the computability of
the set $\{ M  \condition$ there is some input $x$ on which
$M(x)$ does not halt$\}$).

    We mention in passing a related research line 
    about ``independence results in computer science.''
    That line started with work of 
    Hartmanis and Hopcroft~\protect\cite{har-hop:j:ind}
    based on the nontransparency of machines, and has now reached
    the point where it has been shown, by 
    Regan,
    that for each 
    fixed recursively axiomatizable proof 
    system there is a {\em language\/} with 
    certain properties that
    the system cannot prove, no matter how the 
    language is represented 
    in the system 
    (say, by a Turing machine accepting it).
    For instance, 
    for each 
    fixed recursively axiomatizable proof 
    system 
    there is a low-complexity language
    that is infinite, but for no Turing machine
    accepting the language can the proof
    system prove that that 
    Turing machine accepts an infinite language.
    See~\protect\cite{reg:j:index,reg:j:topology} and the 
    references therein.}
Recently, Kari~\cite{kar:j:rice-for-cellular} 
has proven, for cellular automata, an analog of 
Rice's Theorem:  All nontrivial properties of limit sets
of cellular automata are 
undecidable.

A bold and exciting paper of Borchert and
Stephan~\cite{bor-ste:c:rice} proposes and initiates the search for
{\em complexity-theoretic\/} analogs of Rice's Theorem. 
Borchert and Stephan note that Rice's
Theorem deals with properties of programs, and they suggest as a
promising complexity-theoretic analog properties of boolean circuits.
In particular,
they focus on counting properties of circuits. Let $\nats$ denote
$\{0, 1, 2, \ldots \}$. Boolean functions are functions that for some
$n$ map $\{0,1\}^{n}$ to $\{0,1\}$. Circuits built over boolean gates
(and encoded in some standard way---in fact, for
simplicity of expression, we will often 
treat a circuit and its encoding as interchangeable) 
are ways of representing boolean
functions. As Borchert and Stephan point out, the parallel is a close
one.  Programs are concrete objects that correspond in a many-to-one
way with the semantic objects, languages.  Circuits (encoded into
$\sigmastar$) are concrete objects that correspond in a many-to-one
way with the semantic objects, boolean functions. Given an arity $n$
circuit~$c$, $\#(c)$ denotes under how many of the $2^n$ possible
input patterns $c$ evaluates to~1.

\begin{defn}
  \begin{enumerate}
  \item {\rm \cite{bor-ste:c:rice}} \quad Each $A \seq \nats$ is a
    {\em counting property of circuits}. 
    If $A \neq \emptyset$, we say it is
    a nonempty property, and if $A \neq \nats$, we say it is a proper
    property.

  \item {\rm \cite{bor-ste:c:rice}} \quad Let $A$ be a counting
    property of circuits. The {\em counting problem for~$A$},
    $\counting(A)$, is the set of all circuits $c$ such that $\#(c)
    \in A$.
    
  \item {\rm (see~\cite{gar-joh:b:int})} \quad For each complexity
    class ${\calc}$ and each set $B \seq \sigmastar$, we say $B$ is
    {\em ${\calc}$-hard\/} if $(\forall L \in {\calc})\, [L
    \leq_{\rm T}^{\rm p} B]$, where as is standard
    $\leq_{\rm T}^{\rm p}$ denotes polynomial-time 
    Turing reducibility.

  \item {\rm (following usage of~\cite{bor-ste:c:rice})} \quad Let $A$ 
    be a counting property and let ${\calc}$ be a
    complexity class. By convention, we say that counting property $A$
    is {\em ${\calc}$-hard\/} if the counting problem for~$A$,
    $\counting(A)$, is ${\calc}$-hard. (Note in particular that by
    this we do not mean ${\calc} \seq \p^{A}$---we are
    speaking just of the complexity of $A$'s counting problem.)
\end{enumerate}
\end{defn}

For succinctness and naturalness, and as it introduces no ambiguity
here, throughout this paper we use ``counting'' to refer to what
Borchert and Stephan originally referred to as
``absolute counting.''  For completeness, we mention that their sets
$\counting(A)$ are not entirely
new:  For each $A$, $\counting(A)$ is
easily seen (in light of the fact that circuits can be parsimoniously
simulated by Turing machines, which themselves, as per the references
cited in the 
proof of Theorem~\ref{t:dt}, can be parsimoniously transformed into
boolean formulas) to be many-one equivalent to the set,
known in the literature as 
$\sat_A$ or 
\mbox{$A$-$\sat$},
$\{ f \condition\,$the number of satisfying assignments
to boolean formula $f$ is an integer contained in the set 
$A\}$~\cite{gun-wec:j:count,cai-gun-har-hem-sew-wag-wec:j:bh2}.  
Thus,
$\counting(A)$ inherits the various properties that the
earlier papers on $\sat_A$ established for
$\sat_A$,
such as completeness for certain counting 
classes.
We will at times draw on this earlier work to gain insight into
the properties of 
$\counting(A)$.

The results of Borchert and Stephan
that led to the research 
reported on in the present paper are the following.
Note that Theorem~\ref{t:nt} is a 
partial analog of
Theorem~\ref{t:rversiontwo-nowone},\footnote{Passing on 
a comment from an anonymous referee, we mention that the reader 
may want to also compare the 
$\up \bigoplus \coup $ occurrence in Borchert and 
Stephan~\protect\cite{bor-ste:c:rice} with the 
so-called Rice-Shapiro Theorem (see, e.g.,~\cite{rog:b:rft,reg:j:index}).  
We mention that in making such a 
comparison one should keep in mind that the 
Rice-Shapiro Theorem deals with showing {\em non-membership\/}
in RE and coRE, rather than with showing many-one {\em hardness\/} for 
those classes.}
and Corollary~\ref{c:ea} is a partial
analog of Corollary~\ref{cwast:rversionone-nowtwo}. 
$\up\bigoplus\coup$ denotes $\{ A \oplus B \condition
A\in \up \land B \in \coup\}$, where $A\oplus B = 
\{0x \condition x\in A\} \cup \{1y \condition y\in B\}$.

\begin{thm} \label{t:nt}
  ({\rm \cite{bor-ste:c:rice}}, see also the comments at 
the start of the proof of Theorem~\ref{t:dt})
 \quad Let $A$ be a nonempty proper
  subset of~$\nats$. Then one of the following three classes is
  $\leq_{\rm m}^{\rm p}$-reducible to $\counting(A)$: $\np$, $\conp$, or $\up
  \bigoplus \coup$.
\end{thm}

\begin{cor} \label{c:ea}
  ({\rm \cite{bor-ste:c:rice}}, see also the comments at 
the start of the proof of Theorem~\ref{t:dt})
\quad 
  Every nonempty proper counting property of circuits is $\up$-hard.
\end{cor}

Borchert and Stephan's paper proves 
a number of other results---regarding an
artificial existentially quantified circuit type yielding NP-hardness,
definitions and results
about counting properties over rational numbers and
over~$\Z$, and so on---and we highly commend their paper to the
reader.  They also give a very interesting motivation.  
They show that, in light of the work of 
Valiant and Vazirani~\cite{val-vaz:j:np-unique}, 
any nontrivial counting property of circuits is hard
for either NP or coNP, with respect to {\em randomized reductions}.
Their paper and this one seek to find to what extent 
or in what form this behavior 
carries over to deterministic reductions.

The present paper makes the following contributions. First, we extend
the above-stated results 
of Borchert and Stephan, 
Theorem~\ref{t:nt} and Corollary~\ref{c:ea}. Regarding
the latter, from
the same hypothesis as their Corollary~\ref{c:ea} we derive a
stronger lower bound---$\up_{{\bigo }(1)}$-hardness. That is, we
raise their lower bound from {\em unambiguous\/} nondeterminism to
{\em low-ambiguity\/} nondeterminism.  Second, we show that
our improved lower 
bound cannot be further strengthened to
SPP-hardness unless an unlikely complexity class containment---$\spp
\seq \p^{\np}$---occurs. Third, we nonetheless under a very natural
hypothesis raise the lower bound on the hardness of counting
properties to SPP-hardness. The natural hypothesis strengthens the
condition on the counting property to require not merely that it is
nonempty and proper, but also that it is infinite and
coinfinite
in a way that can be certified by polynomial-time
machines. 

\section{The Complexity of Counting Properties of Circuits}

All the notations and definitions in this 
paragraph are standard
in the literature.
Fix the alphabet $\Sigma = \{0,1\}$.
$\fp$
denotes the class of polynomial-time computable functions from
$\sigmastar$ to~$\sigmastar$. 
Given 
any two sets $A, B \seq \sigmastar$,
we say $A$ polynomial-time many-one reduces to $B$ ($A\leq_{\rm m}^{\rm p} B$)
if $(\exists f \in \fp)\, (\forall x \in \sigmastar)\, [x \in A \ 
\Longleftrightarrow \ f(x) \in B]$. 
For each set $A$, 
$||A||$ denotes the number of
elements in~$A$.   The length of each string $x \in \sigmastar$ is
denoted by~$|x|$.  We use DPTM (respectively, NPTM)
as a shorthand for deterministic
polynomial-time Turing machine
(nondeterministic
polynomial-time Turing machine). 
Turing machines 
and their languages (with or without oracles) are denoted
as is standard, as are complexity classes (with or 
without oracles), e.g., 
$M$, $M^A$, $L(M)$, $L(M^A)$, $\p$, and $\p^A$.
We allow both languages and functions
to be used as oracles.  In the latter case, the model
is the standard one, namely, when query $q$ is asked to a function
oracle $f$ the answer is $f(q)$.
For each $k\in \nats$,  the notation
``$[k]$'' denotes a restriction
of at most $k$ oracle questions (in a sequential---i.e., ``adaptive''
or ``Turing''---fashion).  For example,
$\p^{\fp[2]}$ denotes
$\{ L \condition
(\exists~\mbox{DPTM}~M)
(\exists f \in \fp) \,
[ L=L(M^f) \land
(\forall x \in \sigmastar)\, [M^f(x)$ makes at most 
two oracle queries$]]\}$, which happens to be merely an
ungainly way of describing the complexity class P\@.~~The 
notation ``$[\bigo (1) ]$'' denotes that, for some constant
$k$, a ``$[k]$'' restriction holds.

We will define, in a uniform way via counting functions, some
standard ambi\-guity-limited classes and counting classes. 
To do this, we will take the 
standard ``\#'' operator~(\cite{tod:thesis:counting}
for the concept and~\cite{vol:thesis:functions} for the 
notation, see the discussion in~\cite{hem-vol:j:satanic})
and will make it flexible enough to describe 
a variety of types of counting functions that are well-motivated
by existing language classes.
In particular, we will add a general
restriction on the maximum value it can take on.  (For the 
specific case of a polynomial 
restriction such an operator, $\#_{\rm few}$, was 
already introduced by Hemaspaandra
and Vollmer~\cite{hem-vol:j:satanic}, see below).

\begin{defn} \label{d:general-sharp}
For each function $g: \nats \rightarrow \, \nats$
and each class $\calc$, define
$\mbox{$\#_{g} \cdot \calc$}  = \{ f: \sigmastar \rightarrow \, \nats 
\conditionX \allowbreak
(\exists L \in \calc )\, 
\allowbreak (\exists \mbox{\rm\  polynomial }$s$)\, \allowbreak
(\forall x \in \sigmastar)\, \allowbreak [ f(x) \leq g(|x|)  \land 
\allowbreak
|| \{y \condition  \allowbreak |y|=s(|x|) \land  \allowbreak
\langle x,y \rangle \in L \} || = f(x)] 
\}$.
\end{defn}

Note that for the very special case
of $\calc = \p$, which is the case of importance in the 
present paper, this definition simply yields classes
that speak about the number of accepting paths 
of Turing machines that obey some constraint on their 
number of accepting paths.  In particular, the following
clearly holds for each~$g$:
$\mbox{$\#_{g}\cdot \p$} = \{ f: \sigmastar \rightarrow \, \nats 
\condition  \allowbreak
(\exists \mbox{ NPTM } N)\, (\forall x \in \sigmastar)\, [ 
N(x)$ has exactly
$f(x)$
accepting paths and $f(x) \leq g(|x|)] 
\}$.

In using Definition~\ref{d:general-sharp}, we will allow 
a bit of informality regarding describing the functions~$g$.  
For example, we will write $\#_1$ when formally we should 
write $\#_{\lambda x.1}$, and so on in similar cases.
Also, we will now define some versions of the $\#_g$ operator
that focus on collections of bounds of interest to us.

\begin{defn} \label{d:specific-sharp}
\begin{enumerate}
\item
For each class $\calc$,
$\mbox{$\#_{\littleconst }\cdot \calc$}  
= \{ f: \sigmastar \rightarrow \, \nats 
\condition  \allowbreak
(\exists k \in \nats )\,  \allowbreak
[ f \in \mbox{$\#_k \cdot \calc$}]
\}$.

\item  {\rm \cite{hem-vol:j:satanic} } \quad
For each class $\calc$,
$\mbox{$\#_{\rm few }\cdot \calc$}  = \{ f: \sigmastar \rightarrow \, \nats 
\condition  \allowbreak
(\exists \mbox{\rm\ polynomial } s)\,  \allowbreak
[ f \in \mbox{$\#_s \cdot \calc$}]
\}$.
\end{enumerate}
\end{defn}

\begin{defn}
  \begin{enumerate}
  \item {\rm \cite{val:j:enumeration}} \quad
$ \sharpp = \{ f: \sigmastar \rightarrow \, \nats 
\condition  \allowbreak
(\exists \mbox{ NPTM } N)\, \allowbreak
(\forall x \in \sigmastar)\, \allowbreak [ N(x)$
has exactly $f(x)$ accepting paths$] 
\}$.

\item {\rm \cite{val:j:checking}} \quad $\up = \{ L \condition 
 \allowbreak
  (\exists f \in \mbox{$\#_{1}\cdot \p$})\,  \allowbreak
(\forall x \in \sigmastar)\, \allowbreak [x \in
  L \,\Longleftrightarrow\, \allowbreak f(x) > 0 ] \}$.  
  
\item {\rm (\cite{bei:c:up1}, see 
also \cite{wat:j:hardness-one-way})}
\quad For each $k \in \nats - \{0\}$,
$\up_{\leq k} = \{ L \condition 
 \allowbreak (\exists f \in 
\mbox{$\#_{k}\cdot \p$})\, \allowbreak
(\forall x \nolinebreak \in \nolinebreak \sigmastar)\, 
\allowbreak [x \nolinebreak \in \nolinebreak L \,\Longleftrightarrow\, 
\allowbreak
f(x) > 0
] \}$.

\item {\rm (\cite{hem-zim:tOutByJour:balanced}, see 
also~\cite{bei:c:up1})} \quad 
  $\up_{\bigo (1)} = \{ L 
\condition  \allowbreak
  (\exists f \in 
\mbox{$\#_{\littleconst
}\cdot \p$}) \, \allowbreak
  (\forall x \nolinebreak \in 
\nolinebreak \sigmastar)\, \allowbreak [x \in
  L \,\Longleftrightarrow\, \allowbreak  f(x) > 0 ]
\}$.
{\rm{}(}Equivalently, $\up_{\bigo (1)} = \bigcup_{k \geq 1}
\up_{\leq k}$.{\rm{})}

\item {\rm \cite{all-rub:j:print}}
  \quad $\fewp = \{ L 
\condition  \allowbreak
  (\exists f \in 
\mbox{$\#_{\rm few }\cdot \p$}) \,  \allowbreak
  (\forall x \nolinebreak \in 
\nolinebreak \sigmastar)\, \allowbreak [x \in
  L \,\Longleftrightarrow\, \allowbreak f(x) > 0 ] 
\}$.
  
\item {\rm \cite{cai-hem:j:parity}} \quad $\few = 
\p^{(\#_{\rm few }\cdot \p)[1]} $.

\item $\const = 
\p^{(\#_{\littleconst }\cdot \p)[\bigo (1)]}$.\footnote{As we will 
note in the proof of
Theorem~\ref{t:dt},
$\p^{(\#_{\littleconst }\cdot \p)[\bigo (1)]}  = 
\p^{(\#_{\littleconst }\cdot \p)[1]}$.  Thus, the definition
of $\const$ is more analogous to the definition 
of $\few$ than one might realize at first glance.}

\item {\rm \cite{fen-for-kur:j:gap,hem-ogi:j:closure}}
  \quad   $\spp = \{ L
\condition  \allowbreak
(\exists f \in \sharpp)\, \allowbreak (\exists g \in \fp) 
\, \allowbreak
(\forall x \in \sigmastar) \,   \allowbreak
[(x \not\in L \,\Longleftrightarrow\, f(x) = 2^{|g(x)|}) 
\land 
\allowbreak (x \in L \,\Longleftrightarrow\, 
\allowbreak f(x) = 2^{|g(x)|} + 1)]
\}$.
\end{enumerate}
\end{defn}

It is well-known that $\up = \up_{\leq 1} \seq \up_{\leq 2} \seq
\cdots \seq \up_{{\bigo}(1)} \seq \fewp \seq \few \seq \spp$ (the
final containment is due to K\"obler et
al.~\cite{koe-sch-tod-tor:j:few}, see
also~\cite{fen-for-kur:j:gap} for a more general result), and
clearly 
$\up_{{\bigo}(1)} \seq \const \seq \few$.
$\spp$ plays a central role in much of complexity
theory (see~\cite{for:b:counting-survey}), and in 
particular is closely linked to 
the closure properties
of $\sharpp$~\cite{hem-ogi:j:closure}.
Regarding relationships with the
polynomial hierarchy, $\p \seq \up \seq \fewp \seq \np$, and $\few
\seq \p^{\fewp}$ (so $\few \seq \p^{\np}$). It is widely suspected
that $\spp \not\seq \ph$ (where PH denotes the polynomial hierarchy),
though this is an open research question.
UP, $\up_{{\bigo }(1)}$, and $\fewp$ are tightly connected to the
issue of whether one-way functions
exist~\cite{gro-sel:j:complexity-measures,all-rub:j:print,hem-zim:tOutByJour:balanced},
and Watanabe~\cite{wat:j:hardness-one-way} has shown that $\p = \up$
if and only if $\p = \up_{{\bigo }(1)}$.

Intuitively, UP captures the notion of unambiguous nondeterminism,
FewP allows polynomially ambiguous nondeterminism and, most relevant
for the purposes of the present paper, $\up_{{\bigo }(1)}$
allows constant-ambiguity nondeterminism.
Corollary~\ref{c:cg} raises the  UP lower bound
of Borchert and Stephan
(Corollary~\ref{c:ea}) to a $\up_{{\bigo }(1)}$ lower bound. This is
obtained via the even stronger bound provided by Theorem~\ref{t:dt},
which itself extends Theorem~\ref{t:nt}. 

\begin{thm}
\label{t:dt}
Let $A$ be a nonempty proper subset of~$\nats$. Then one of the
following three classes is $\leq_{\rm m}^{\rm p}$-reducible to 
$\counting(A)${\rm :}
$\np$, $\conp$, or $\const$.
\end{thm}

\begin{cor}
  \label{c:cg} 
  Every nonempty proper counting property of circuits is $\up_{{
      O}(1)}$-hard 
(indeed, is even
$\up_{{O}(1)}$-$\leq_{\rm 1\hbox{-}tt}^{\rm p}$-hard\/\footnote{Where
$\leq_{\rm 1\hbox{-}tt}^{\rm p}$ as is standard 
denotes polynomial-time 1-truth-table
reductions~\cite{lad-lyn-sel:j:com}.}$\!$).
\end{cor}

Our proof applies a constant-setting technique that Cai and
Hemaspaandra (then Hemachandra)~\cite{cai-hem:j:parity}
used to
prove that $\fewp \seq \parityp$, and that 
K\"obler et al.~\cite{koe-sch-tod-tor:j:few}
extended to show that $\few \subseteq \spp$.  Borchert, 
Hemaspaandra, and 
Rothe~\cite{bor-hem-rot:t:powers-of-two}
have used the method to study the complexity of 
equivalence problems 
for OBDDs (ordered binary decision diagrams) and other 
structures.

{\bf Proof of Theorem~\protect\ref{t:dt}.} 
Let $A$ be a nonempty proper
subset of~$\nats$. 
The paper of Borchert and Stephan~\cite{bor-ste:c:rice} (see
Theorem~\ref{t:nt} above) 
and---using different nomenclature---earlier 
papers~\cite{gun-wec:j:count,cai-gun-har-hem-sew-wag-wec:j:bh2}~have 
shown that
(a)~if $A$ is finite and nonempty, then
$\counting(A)$ is $\leq_{\rm m}^{\rm p}$-hard for coNP, and (b)~if $A$ is
cofinite and a proper subset of $\nats$, 
then $\counting(A)$ is $\leq_{\rm m}^{\rm p}$-hard for NP. 

We will now show that 
if $A$ is infinite and
coinfinite, then $\counting(A)$ is $\leq_{\rm m}^{\rm p}$-hard for
$\const$.
Actually, it is not hard to see that
$\p^{(\#_{\littleconst}\cdot \p)[\bigo(1)]} = \p^{(\#_{\littleconst}\cdot
  \p)[1]}$, 
and so we need
deal just with 
$\p^{(\#_{\littleconst}\cdot 
\p)[1]}$.\footnote{This is a property that 
seems to be deeply dependent on the ``const''-ness.  For example,
it is not known whether $\p^{\sharpp[1]} = \p^{\sharpp[2]}$,
and indeed it is known that if this seemingly 
unlikely equality holds then two complexity classes
associated with self-specifying machines
are equal~\protect\cite{hem-hem-wec:t2:self-specifying}.}
The reason the just-mentioned equality
holds is that since each of the constant number of questions (say $v$)
has at most a constant number of possible answers (say $w$) 
one can 
by brute force accept each
$\p^{(\#_{w}\cdot \p)[v]}$ 
language via DPTMs that make at most
${\frac{w^{v+1}-1}{w-1}}$
queries in a truth-table fashion to a
function (in fact, the same function) from $\#_{w} \cdot \p$.
However, the same encoding argument 
(\cite{cai-hem:j:parity}, see 
also~\cite{pap-zac:c:two-remarks})
that shows that
bounded-truth-table access to a $\sharpp$ function can be replaced by
one query to a $\sharpp$ function in fact also shows 
that 
${\frac{w^{v+1}-1}{w-1}}$-truth-table 
access to a $\#_{w} \cdot \p$
function can be replaced by one query to a 
$\#_{w^{ {\frac{w^{v+1}-1}{w-1}} } - 1 } \cdot \p$ 
function.

Let $B$ be an arbitrary set in
$\p^{(\#_{\littleconst}\cdot \p)[1]}$, 
and let $B\in \p^{(\#_{\littleconst}\cdot \p)[1]}$
be witnessed by some DPTM
$M$ that makes at most one query (and without loss 
of generality we assume that on each input $x$ it in 
fact makes {\em exactly\/} one query) to
some function $h \in \#_{\littleconst}\cdot \p$. Let $N'$ be
some NPTM and let $k$ be some constant such that for each string $z
\in \sigmastar$, $N'(z)$ has exactly $h(z)$ accepting paths and $h(z)
\leq k$. Such a machine exists by the equality mentioned 
just after Definition~\ref{d:general-sharp}.
For each input $x$ to $M$, let $q_x$ be the single query
to $h$ in the run of $M(x)$.
We will call a nonnegative integer $\ell$ such that 
$\ell \in A$ and $\ell +1 \not\in A$ a
{\em boundary event (of $A$)}, and we will
in such cases call $\ell + 1$ a 
{\em boundary shadow}
(see, for comparison,~\cite{gol:thesis:nt,gol-jos-you:c:internal-structure,gol-hem-jos-you:j:nt}).
Since $A$ is infinite and coinfinite, note that it 
has infinitely many boundary events.
We now define a function $g \in \sharpp$ such that 
\begin{eqnarray}
(\forall x \in \sigmastar)
[M^{h}(x) 
\mbox{ accepts} \,\, \Longleftrightarrow \, g(x) \in A]. \label{eqn:iff}
\end{eqnarray}
We will do so by mapping $x$ for which $M^{h}(x)$ accepts to 
boundary events, and by mapping $x$ for which 
$M^{h}(x)$ rejects to boundary shadows.
To define~$g$, we
now describe an NPTM $N$ that witnesses $g \in \sharpp$. 

On input~$x$,
$N$ first computes 
the oracle query $q_x$ of~$M(x)$. Then $N(x)$ chooses
$k+1$ constants $c_0, c_1, \ldots , c_k$ as follows.

$M^{\lambda z . j }(x) \in \{0,1\}$ denotes the result of the
computation of $M(x)$ assuming the answer of the oracle was $h(q_x) =
j$, where our convention is that 
$M^{\lambda z . j }(x) = 0$ stands
for ``reject'' and $M^{\lambda z . j }(x) = 1$ stands for ``accept.''
Let $a_0$ be the least boundary event
of $A$ (recall that boundary events are nonnegative
integers, and thus it does make sense to speak of the 
least
boundary event).
Initially, choose
\[
c_0 = \left\{ 
\begin{array}{ll}
a_0     & \mbox{if $M^{\lambda z . 0 }(x) = 1$} \\
a_0 + 1 & \mbox{if $M^{\lambda z . 0 }(x) = 0$.}
\end{array}
\right.
\]
Successively, for $i = 1, \ldots , k$, do the following:
\begin{itemize}
\item Let $c_0, \ldots , c_{i-1}$ be the constants that have already
  been chosen.  For each $i \in \nats$, $\binom{i}{0} = 1$
as is standard.  Let $b_i =
\binom{i}{0} c_0 + \binom{i}{1} c_1 + \binom{i}{2} c_2 + 
\cdots + \binom{i}{i-1} c_{i-1}$.

\item Let $a_i$ be the least boundary event of $A$
  such that $b_i \leq a_i$.

\item Set the constant
\[
c_i = \left\{ 
\begin{array}{ll}
a_i - b_i     & \mbox{if $M^{\lambda z . i }(x) = 1$} \\
a_i + 1 - b_i & \mbox{if $M^{\lambda z . i }(x) = 0$.}
\end{array}
\right.
\]
\end{itemize}

After having chosen these constants,\footnote{Note that 
as $2^{k+1}$ is also a constant we could alternatively
simply build into the machine $N$ a table that, for each of 
the $2^{k+1}$ behavior patterns $M$ can have on an input 
(in terms of whether it accepts or rejects for each
given possible answer from the oracle), states what 
constants $c_0, \ldots ,c_k$ to use.  The procedure just 
given would be used to decide the values of this 
table, which would then be hardwired into $N$.}
$N(x)$ guesses an integer $j \in
\{0, 1, ...  , k\}$, and immediately splits into $c_0$ accepting
paths if the guess was $j = 0$. For each $j > 0$ guessed, $N(x)$
nondeterministically guesses each $j$-tuple of 
distinct paths of~$N'(q_x)$. On
each such path of~$N(x)$, where the $j$-tuple $(\alpha_1, \alpha_2,
\ldots, \alpha_j)$ of paths of $N'(q_x)$ has been guessed,
$N(x)$ splits into exactly $c_j$ accepting paths if each~$\alpha_m$,
$1 \leq m \leq j$, is an accepting path of~$N'(q_x)$. If, however, for
some $1 \leq m \leq j$, $\alpha_m$ is a rejecting path of~$N'(q_x)$,
then $N(x)$ simply rejects (along the current 
path). This completes the description of~$N$.

Recall that $h(q_x) 
\in \{0, 1, \ldots , k \}$
is the true answer of the oracle. Then, by the above construction, the
number of accepting paths of $N(x)$ is
$$g(x) = c_0 + \binom{h(q_x) }{1} c_1 + \binom{h(q_x) }{2} c_2 + 
\cdots
+ 
\binom{h(q_x) }{h(q_x)  - 1} 
c_{h(q_x)  - 1} + \binom{h(q_x) }{h(q_x) } c_{h(q_x) }.$$
However, $c_{h(q_x) }$ 
has been chosen such that $g(x) = b_{h(q_x) } + c_{h(q_x) } =
a_{h(q_x) } \in A$ if $M^{h}(x)$ accepts, and 
$g(x) = b_{h(q_x) } + c_{h(q_x) } =
a_{h(q_x) } + 1 \not\in A$ if $M^{h}(x)$ rejects. 
Since each $a_{i}$, $0\leq i \leq k$, is a boundary event and each
$a_{i}+1$, $0\leq i \leq k$, is a boundary shadow,
this completes our proof of 
Equation~\ref{eqn:iff}.

By the well-known observation (mentioned by Garey and 
Johnson~\cite[p.~169]{gar-joh:b:int}, see also the
primary sources~\cite{sim:thesis:complexity,val:j:enumeration}) that the
many-one reductions 
of the
Cook-Karp-Levin Theorem can be altered so as to be ``parsimonious,''
there is a
$\leq_{\rm m}^{\rm p}$-reduction that 
on input $x$ ($N$ is not an input to this
$\leq_{\rm m}^{\rm p}$-reduction, but rather is hardwired
into the reduction)
outputs
a boolean formula $\phi_{x}(y_1, ... , y_n)$, where
$n$ is polynomial in~$|x|$, such that the number of satisfying
assignments of $\phi_{x}(y_1, \ldots , y_n)$ equals~$g(x)$. Let
$c_{\phi_{x}}(y_1, \ldots , y_n)$ denote 
(the representation of) a circuit for
that formula.  There is a DPTM implementing 
this formula-to-circuit transformation.
Our reduction from $B$ to $\counting(A)$ is defined by
$f(x) = c_{\phi_{x}}(y_1, \ldots , y_n)$. Clearly, $f$ is
polynomial-time computable, which together with 
Equation~\ref{eqn:iff}
implies $B \leq_{\rm m}^{\rm p} \counting(A)$ via~$f$.~\qed

Corollary~\ref{c:cg} raised the lower bound of Corollary~\ref{c:ea}
{}from UP to~$\up_{{\bigo }(1)}$. It is natural to wonder whether the
lower bound can be raised to~SPP\@.  This is 
especially true in light of the
fact that Borchert and Stephan obtained SPP-hardness results
for their notions of ``counting problems over $\Z$'' and 
``counting problems over the
rationals''; their UP-hardness result for standard counting problems
(i.e., over~$\nats$) is the short leg of their paper.  However, we
note that extending the hardness lower bound to SPP under the same
hypothesis seems unlikely. 
Let~BH denote the boolean
hierarchy~\cite{cai-gun-har-hem-sew-wag-wec:j:bh1}.  
It is well-known that $\np
\seq \bh \seq \p^{\np} \seq \ph$.
\begin{prop} \label{p-was-t:hz}
  If $A \seq \nats$ is finite or cofinite, then $\counting(A) \in \bh$.
\end{prop}

This result needs no proof, as it follows easily
from
Lemma~3.1 and Theo\-rem~3.1.1(a) of~\cite{cai-gun-har-hem-sew-wag-wec:j:bh2} 
(those results
exclude the case $0\in A$ but their proofs clearly apply also to that
case)
or from~\cite[Theorem~15]{gun-wec:j:count}, 
in light of the relationship between $\counting(A)$ and
$\sat_A$ mentioned earlier in the present paper.
Similarly, from earlier work one can
conclude that, though for all finite and cofinite 
$A$ it holds that $\counting(A)$ is in the boolean
hierarchy, these problems are not good candidates 
for complete sets for that hierarchy's higher levels---or even
its second level.
In particular, from the approach of the 
theorem and proof 
of~\cite[Theorem~3.1.2]{cai-gun-har-hem-sew-wag-wec:j:bh2}
(see also~\cite[Theorem~15]{gun-wec:j:count})
it is not too hard to see that 
$(\exists B)\,[ (\forall \mbox{ finite } A)\, 
[\counting(A)$ is not $\leq_{\rm m}^{ {\rm p},B}$-hard 
for $\np^B] \land
(\forall \mbox{ cofinite } A)\, 
[\counting(A)$ is not $\leq_{\rm m}^{ {\rm p},B}$-hard 
for $\conp^B]]$.

In light of the fact that SPP-hardness means
SPP-$\leq_{\rm T}^{\rm p}$-hardness, 
the bound of Proposition~\ref{p-was-t:hz}
yields the following result (one can equally well
state the stronger claim that 
no finite or cofinite counting property of circuits
is $\spp$-$\manyone$-hard unless $\spp \seq \bh$).

\begin{cor} \label{c:vp}
  No finite or cofinite counting property of circuits
  is $\spp$-hard unless $\spp \seq \p^{\np}$.
\end{cor}

Though we have not in this paper discussed models of relativized
circuits and 
relativized formulas to allow this work to relativize cleanly
(and we do not view this as an important issue), we mention in passing
that there is a relativization in which SPP is not contained in
$\p^{\np}$ (indeed, relative to which SPP strictly contains the
polynomial hierarchy)~\cite{for:b:counting-survey}.

Corollary~\ref{c:vp} makes it clear that if we seek to prove the
SPP-hardness of counting properties, we must focus only on counting
properties that are simultaneously infinite and coinfinite. Even this
does not seem sufficient. The problem is that there are infinite,
coinfinite sets having ``gaps'' so huge as to make
the sets  have seemingly no interesting usefulness at many lengths
(consider, e.g., the set $\{i \condition
(\exists j)\, [ i = {\tt AckermannFunction}(j,j)] \}$). Of course, in
a recursion-theoretic context this would be no problem, as a Turing
machine in the recursion-theoretic world is free from time constraints
and can simply run until it finds the desired structure (which we
will
see is a boundary event). 
However, in
the world of complexity theory we operate within (polynomial) time
constraints. Thus, we consider it natural to add a hypothesis, in our
search for an SPP-hardness result, requiring that infiniteness and
coinfiniteness of a counting property be constructible in a
polynomial-time manner.

Recall that a set of nonnegative 
integers is infinite exactly if it has no largest element. We will
say that a set is $\p$-constructibly infinite if there is a
polynomial-time function that yields elements of the set at least as 
long as each given input.

\begin{defn}\label{d:construc-changed}
\begin{enumerate}
\item  \label{p:defpartone}
Let $B \seq \sigmastar$. We say that 
$B$ is {\em$\p$-constructibly
    infinite\/} 
if 
\[
(\exists f \in \fp)\, (\forall x \in \sigmastar)\, [f(x) \in B 
\land |f(x)| \geq |x| ].
\]

\item \label{d:parttwo}
Let us adopt the standard bijection between $\sigmastar$ and
$\nats$---the natural number $i$ corresponds to the lexicographically
$(i+1)$st string in $\sigmastar$: $0 \leftrightarrow \epsilon$, $1
\leftrightarrow 0$, $2 \leftrightarrow 1$, $3 \leftrightarrow 00$,
etc. 
If $A \seq \nats$, we say that $A$ is $\p$-constructibly infinite
if $A$, viewed as a subset of $\sigmastar$ via this bijection, is
$\p$-constructibly infinite according
to Part~\ref{p:defpartone} of this definition.

\item If $A \subseteq \sigmastar$ and $\overline{A}$ 
(or $A \subseteq \nats$ and $\nats - A$)
are $\p$-constructibly
infinite, we will say that 
$A$ is {\em$\p$-constructibly
bi-infinite}.
\end{enumerate}
\end{defn}

The above is our formal, type-correct definition,
and is the definition we employ within our proof of
Theorem~\ref{t:bi-infinite}.  However, Part~\ref{d:parttwo} 
of the definition is a bit
long.
Following a referee's suggestion, as an aside 
we mention
a different, more intuitive definition that happens to yield
the same class.
Let us say that 
$A \seq \nats$ belongs to NICE
if there is a polynomial-time function $g$  (mapping from 
$\nats$ to $\nats$) such that
for all $n \in \nats$ we have $g(n) \in A$ and 
$|g(n)| > |n|$, where both the
explicit ``length-of''s ($|g(n)|$ and $|n|$) 
and the one implicit in speaking of a ``polynomial-time function''
are with respect
to the standard way of writing integers in binary without
superfluous leading zeros.  Though this definition differs from
that of Part~\ref{d:parttwo} of Definition~\ref{d:construc-changed}
(e.g., the boundaries between lengths fall at different 
places),
it in fact is not too hard to see
that it does define exactly the same class;  that
is, NICE is exactly $\{ A \seq \nats \condition
A$ is
$\p$-constructibly infinite$\}$.

Note that some languages that are infinite (respectively, bi-infinite)
are not P-con\-structibly infinite (respectively, bi-infinite), e.g.,
languages with huge gaps between successive elements.

Borchert and Stephan~\cite{bor-ste:c:rice} 
also study ``counting problems over the
rationals,'' and in this 
study they use a root-finding-search approach to establishing lower
bounds. In the following proof, we apply this type of approach
(by which we mean the successive interval contraction of the same
flavor used when trying to capture the root of a function on
$[\hspace*{1pt}a,\hspace*{1pt} b\hspace*{1pt}]$ when 
one knows initially that, say,  $f(a) > 0$ and
$f(b)< 0$) to counting problems (over~$\nats$). 
In particular, we use the
$\p$-constructibly bi-infinite hypothesis to ``trap'' a
boundary event of~$\overline{A}$.

\begin{thm}\label{t:bi-infinite}
Every $\p$-constructibly bi-infinite counting property 
of circuits is $\spp$-hard.
\end{thm}
{\bf Proof.}
Let $A \seq \nats$ be any 
$\p$-constructibly bi-infinite counting property 
of circuits. Let $L$ be any set in 
SPP\@.~~Since $L\in \spp$, there are 
functions $f \in \sharpp$ and $g \in \fp$ 
such that, for each $x \in
\sigmastar$:
$(x \in L  \Longleftrightarrow f(x) = 2^{|g(x)|} + 1)  \land  
(x \not\in L  \Longleftrightarrow f(x) = 2^{|g(x)|})$.
Let $h$ and $\overline{h}$ be FP functions certifying
that $A$ and $\overline{A}$ are $\p$-constructibly infinite, in
the exact sense of Part~\ref{d:parttwo} of 
Definition~\ref{d:construc-changed}. We will
describe a DPTM $N$ that
$\leq_{\rm T}^{\rm p}$-reduces $L$ to $\counting(A)$.  For clarity, let
$\widehat{w}$ henceforth denote the natural number that in the above
bijection between $\nats$ and $\sigmastar$ corresponds to the
string~$w$.
For
convenience, we will sometimes view $A$ as a subset of $\nats$ and
sometimes as a subset of $\sigmastar$ (and in the latter case 
we implicitly mean the transformation of $A$ to strings 
under the above-mentioned bijection).

Since clearly $A \leq_{\rm m}^{\rm p} \counting(A)$,\footnote{Either 
one can encode a string $n$ (corresponding
  to the number $\widehat{n}$ in binary) directly into a circuit
  $c_n$ such that $\#(c_n) = \widehat{n}$ (which is easy to do), 
  or one can note the following indirect transformation: Let $N'$
  be an NPTM that on input $n$ produces exactly $\widehat{n}$
  accepting paths.  Using a parsimonious
  Cook-Karp-Levin reduction (as described earlier), we easily obtain a
  family of circuits $\{\widetilde{c}_n\}_{n\in\sigmastar}$ 
  such that, for each $n\in \sigmastar$, $\#(\widetilde{c}_n) = 
   \widehat{n}$.  
  }
we for convenience will sometimes informally speak as if 
the set $A$ (viewed via the bijection as a subset of 
$\sigmastar$) is an oracle of the reduction.
Formally, when we do so, this should be viewed as a shorthand for the
complete $\leq_{\rm T}^{\rm p}$-reduction that consists of the
$\leq_{\rm T}^{\rm p}$-reduction between $L$ and $A$ followed by the
$\leq_{\rm m}^{\rm p}$-reduction between $A$ and $\counting(A)$.

We now describe the machine~$N$. On input~$x$, $|x| = n$,
$N$ proceeds in three steps.  (As a shorthand, we will 
consider $x$ fixed and will write $N$ rather than
$N^{\mbox{\scriptsize\tt Counting}(A)}(x)$.)

{\bf (1)}\quad  $N$ runs $\overline{h}$ and $h$ 
  on suitable inputs to find
  certain sufficiently large strings in $\overline{A}$ and~$A$. In
  particular, let $\overline{h}(0^{|g(x)|+1}) = y$.
  So we have $y \not\in A$ 
  and $|y| \geq |g(x)|+1$, and
  thus $\widehat{y} \geq 2^{|g(x)|+1} - 1 \geq 2^{|g(x)|}$.
  Recall that $|x|=n$.   Since both
  $\overline{h}$ and $g$ are in~FP, there exists a polynomial $p$ such
  that $|y| \leq p(n)$, and thus 
  certainly $\widehat{y} < 2^{p(n)+1}$. So
  let $h(0^{p(n)+2}) = z$, which implies $z \in A$ and $|z| \geq
  p(n)+2$. Thus, 
$\widehat{z} \geq 2^{p(n)+2} - 1 > 2^{p(n)+1} > \widehat{y}$.
Since $h \in \fp$, there clearly exists a polynomial $q$ such that
$\widehat{z} < 2^{q(n)}$. To summarize, $N$ has found in time
polynomial in~$|x|$ two strings $y \not\in A$ and $z \in A$ such that
$2^{|g(x)|} \leq \widehat{y} < \widehat{z} < 2^{q(n)}$.

{\bf (2)} \quad $N$ performs a search on the interval
  $[\hspace*{1pt} \widehat{y} ,\hspace*{1pt} \widehat{z}\hspace*{1pt}
  ] \seq \nats$ to find some $\widehat{u} \in \nats$ 
  that is a boundary event of $\overline{A}$.  That is, $\widehat{u}$
  will satisfy: (a)~$\widehat{y} \leq \widehat{u} \leq \widehat{z}$,
  (b)~$\widehat{u} \not\in A$, and (c)~$\widehat{u} + 1 \in A$. Since
  $\widehat{z} < 2^{q(n)}$, 
the search will terminate in time polynomial
in~$|x|$.  
For completeness
we mention the very standard algorithm to
search to find a boundary event
of $\overline{A}$ (recall the comment above regarding
access to $A$ being in effect available to the algorithm):
\begin{construction}  
\item $\begin{array}{ll}
    \mbox{\bf Input}  & \mbox{$\widehat{y}$ and 
      $\widehat{z}$ satisfying 
$\widehat{y} < \widehat{z}$,
$\widehat{y} \not\in A$, and 
      $\widehat{z} \in A$.} \\
    \mbox{\bf Output} & \mbox{$\widehat{u}$, a boundary event 
      of~$\overline{A}$ satisfying $\widehat{y}\leq
      \widehat{u}\leq\widehat{z}$.}
        \end{array}$
  \begin{block}
      \item $\widehat{u} := \widehat{y}$;
      \item {\bf while } $\widehat{z} > \widehat{u} + 1$ {\bf do }
        \begin{block} 
        \item $\widehat{a} := \lfloor \frac{\widehat{u} +
            \widehat{z}}{2} \rfloor$;
{\bf if } $\widehat{a} \not\in A$
          {\bf  then } $\widehat{u} := \widehat{a}$
          {\bf  else } $\widehat{z} := \widehat{a}$
      \end{block} 
      \item {\bf end while}
  \end{block}
\end{construction}

{\bf (3)}\quad  Now consider the $\sharpp$ function
$
e(\pairs{m,x}) = m + f(x)
$
and the underlying NPTM $E$ witnessing that $e \in \sharpp$.  
Let $d_E$ be the 
parsimonious Cook-Karp-Levin reduction that on each input 
$\pairs{m,x}$ outputs a circuit (representation)
$\widetilde{c}_{\pairs{m,x}}$
such that $\#(
\widetilde{c}_{\pairs{m,x}}) =
e(\pairs{m,x})$. 
Recall that $N$ 
has already computed $\widehat{u}$ (which itself 
depends on $x$ and the oracle).  
$N$, using $d_E$ to build its
query, now queries its oracle, $\counting(A)$, as to 
whether 
$\widetilde{c}_{\pairs{ \widehat{u} - 2^{|g(x)|},x}} \in \counting(A)$, 
and $N$ accepts its input $x$ if and only if the
answer is ``yes.''  
This completes the description of~$N$.

As argued above, $N$ runs in polynomial time. We have to show that it
correctly $\leq_{\rm T}^{\rm p}$-reduces $L$ to $\counting(A)$. Assume $x
\not\in L$. Then $f(x) = 2^{|g(x)|}$, and thus 
$$
e(\pairs{ \widehat{u} - 2^{|g(x)|},x}) = \widehat{u}
\not\in A.
$$ 
This implies that the answer to the query 
``$\widetilde{c}_{\pairs{ \widehat{u} - 2^{|g(x)|},x}} \in 
\counting(A)$?''~is ``no,''
and so $N$ rejects~$x$. 
Analogously, if $x \in L$, then $f(x) =
2^{|g(x)|} + 1$, and thus 
$$
e(\pairs{ \widehat{u} - 2^{|g(x)|},x}) =
\widehat{u} + 1 \in A,
$$ 
and so $N$
accepts~$x$.~\qed

Finally, though we have stressed 
ways in which hypotheses that we feel are natural yield
hardness results, we mention that for a large 
variety of complexity classes 
(amongst them~R, coR, BPP, PP, and FewP)
one can state somewhat 
artificial 
hypotheses for $A$ that ensure that $\counting(A)$ is 
many-one hard for the given class.  For example,   
if $A$ is any set such that either 
$\{ i \condition 
i$ is a boundary event of $A\}$ is P-constructibly 
infinite or 
$\{ i \condition 
i$ is a boundary event of $\overline{A}\}$
is P-constructibly infinite, then $\counting(A)$
is 
\mbox{$\spp$-$\leq_{\rm m}^{\rm p}$-hard}.

\medskip

\noindent {\bf Acknowledgments:}
We are very grateful to 
Bernd Borchert, Edith Hemaspaandra, and Gerd Wechsung for helpful 
discussions and suggestions,
to the anonymous
referees for helpful suggestions,
and to
Lance Fortnow,
Kenneth Regan, and Heribert Vollmer 
for helpful literature pointers and history.
We thank Juris Hartmanis for 
commending to us the importance of finding complexity-theoretic
analogs of index sets, and we commend to the reader, as 
Juris Hartmanis did to us, the open 
issue of finding a crisp complexity-theoretic
analog of the recursion-theoretic work of Hartmanis 
and Lewis~\cite{har-lew:j:undecidable}.

\bibliography{gry}

\showhyphens{ambiguity}
\showhyphens{theorem}
\end{document}